\documentclass[pre, twocolumn, amsmath, amssymb, superscriptaddress]{revtex4}

\usepackage{graphicx}
\usepackage{dcolumn}
\usepackage{bm}
\usepackage{braket}
\usepackage{xcolor}

\usepackage{amsmath}

\newcommand{\beq}{\begin{equation}}
\newcommand{\eeq}{\end{equation}}

\begin{document}

\title{Compact assessment of molecular surface complementarities  enhances neural network-aided prediction of key binding residues}

\author{Greta Grassmann}
\affiliation{Department of Biochemical Sciences ``Alessandro Rossi Fanelli", Sapienza University of Rome, P.Le A. Moro 5, 00185, Rome, Italy}
\affiliation{Center for Life Nano \& Neuro Science, Istituto Italiano di Tecnologia, Viale Regina Elena 291,  00161, Rome, Italy}

\author{Lorenzo Di Rienzo}
\affiliation{Center for Life Nano \& Neuro Science, Istituto Italiano di Tecnologia, Viale Regina Elena 291,  00161, Rome, Italy}

\author{Giancarlo Ruocco}
\affiliation{Center for Life Nano \& Neuro Science, Istituto Italiano di Tecnologia, Viale Regina Elena 291,  00161, Rome, Italy}
\affiliation{Department of Physics, Sapienza University, Piazzale Aldo Moro 5, 00185, Rome, Italy}

\author{Mattia Miotto  \footnote{\label{corr} For correspondence write to: mattia.miotto@roma1.infn.it and edoardo.milanetti@uniroma1.it}
}
\affiliation{Center for Life Nano \& Neuro Science, Istituto Italiano di Tecnologia, Viale Regina Elena 291,  00161, Rome, Italy}

\author{Edoardo Milanetti$^*$}
\affiliation{Center for Life Nano \& Neuro Science, Istituto Italiano di Tecnologia, Viale Regina Elena 291,  00161, Rome, Italy}
\affiliation{Department of Physics, Sapienza University, Piazzale Aldo Moro 5, 00185, Rome, Italy}

\begin{abstract}

Predicting interactions between biomolecules, such as protein-protein complexes, remains a challenging problem. Despite the many advancements done so far, the performances of docking protocols are deeply dependent on their capability of identify binding regions. 
In this context, we present a novel approach that builds upon our previous  works modeling protein surface patches via sets of orthogonal polynomials to identify regions of high shape/electrostatic complementarity. 
By incorporating another key binding property, such as the balance between hydrophilic and hydrophobic contributions, we define new binding matrices that serve an effective inputs for training a neural network.
Our approach also allows for the quantitative definition of a typical binding site area - approximately 10\AA~in radius - where hydrophobic contribution and shape complementarity, which reflects the Lennard-Jones interaction, are maximized.\\
Using this new architecture, CIRNet (Core Interacting Residues Network), we achieve an accuracy of approximately 0.82 in identifying pairs of core interacting residues on a balanced dataset. In a blind search for core interacting residues, CIRNet distinguishes these from decoys with a ROC AUC of 0.72.
This protocol can enahnce docking algorithms by rescaling the proposed poses. When applied to the top ten models from three popular docking server, CIRNet improves docking outcomes, reducing the the average RMSD between the refined poses and the native state  by up to 58\%.

\end{abstract}

\maketitle

\section{Introduction}

\thispagestyle{empty}

\section*{Introduction}
Over 80\% of proteins operate in molecular complexes \cite{rao2014protein}, making the understanding of protein complex formation fundamental for unveiling physiological and pathological cellular processes. Investigating the mechanisms driving protein binding is therefore a crucial challenge in molecular biology, with significant implications for therapeutic and biotechnological applications.\\
While experimental techniques such as X-ray crystallography and NMR spectroscopy allow for rapid and large-scale detection of these complexes, they are often expensive and time-consuming. Additionally, many biological functions rely on transient complexes that are challenging to determine experimentally, even when the structures of the binding partners are known. Computational methods offer a cost-effective alternative and also enable the observation of structural and energetic variations at the atomic level over time. Understanding these variations is essential for predicting protein interactions, particularly in the crowded environment of the cellular interior where they happen.
The cellular environment is characterized by a high concentration of molecules and other components \cite{minton1981excluded, minton1981effect}, with biomolecules at concentrations ranging from 100 to 450 g/L, making up approximately 40\% of the cytoplasm \cite{ellis2001macromolecular, ellis2003join, zimmerman1991estimation}. 
Binding partners have to find themselves among thousands of other types of molecules in the cell and avoid other non-specific interactions. For this to happen, interactions between specific partners have to be precisely and finely tuned \cite{Sheinerman2002}. Accurate prediction and understanding of these interactions require unveiling all components of the binding mechanisms, highlighting the importance of computational methods in this domain \cite{grassmann2024computational}.\\
The introduction of AlphaFold2 \cite{jumper2021highly} in 2021 revolutionized this field by providing 3D structures of proteins (with a resolution close to the experimental data) based solely on their amino acid sequences. This development has significantly enhanced the role of structural information in protein interaction modeling. Nonetheless, competitions that evaluate the latest protein-protein docking algorithms, such as CAPRI \cite{janin2003capri}, continue to seek effective methods for interface and pose prediction. These methods usually produce a high number of models, but they are still lacking reliable scoring functions to discriminate the model closest to the native structure.\\
Docking tools can be classified as direct or template-based. Template-based methods use homology modeling to predict the complex structure by identifying common patterns through multiple sequence alignments (MSA), given that interacting pairs with over 30\% sequence identity often interact similarly \cite{aloy2003relationship}. One of the latest template-based protocols is AlphaFold3, introduced by Abramson \textit{et al.} \cite{abramson2024accurate}. AlphaFold3 can predict complexes including proteins, nucleic acids, small molecules, ions, and modified residues. Thanks to a diffusion-based architecture, it reaches a higher accuracy compared to previous docking tools \cite{abramson2024accurate,yang2023alphafold2,evans2021protein}. However, the availability of homologous proteins is not always guaranteed.
Direct docking methods, which search for the complex structure that minimizes free energy within the conformational space, are not constrained by the availability of homologous proteins. On the other hand, these methods require the definition of a computationally feasible free-energy evaluation model and effective minimization algorithms \cite{vajda2009convergence}. Protein interactions are influenced by a variety of factors, including van der Waals forces, electrostatic interactions, hydrogen bonding, hydrophobic effects, and solvent interactions. Binding sites exhibit a combination of geometric and chemical complementarities, which determine complex formation specificity and binding stability \cite{Gabb1997, Desantis2022, skrabanek2008computational, van1994protein, Miotto_2018, Miotto2020diff, DiRienzo2021boffi, Di_Rienzo2021-yp, Miotto2022server, miotto2023differences,di2023dynamical}. However, incorporating more features increases the computational cost. Given the already substantial cost due to numerous possible contact patches, developing efficient methods that focus on key interaction features is essential.\\
One well-known characteristic of binding interfaces is shape complementarity, which is determined by side-chain rearrangement minimizing van der Waals interactions. In 2021, we proposed a method \cite{Milanetti2021} to evaluate this complementarity using Zernike polynomials to describe the shape of surface portions. This approach, which does not require structural alignment between interacting molecules, leverages the rotational invariance of Zernike polynomials for quick  evaluation of patches compatibility (in terms of Euclidean distance between Zernike vectors) without considering their mutual orientation  \cite{Milanetti2021Insilico, latto, bo2020exploring, biom11121905, Grassmann2022}. This protocol discriminates between transient and permanent protein complexes with an Area Under the Receiver Operating Characteristic Curve (ROC AUC) of approximately 0.8, and blindly identifies true interacting regions in approximately 60\% of the cases \cite{Milanetti2021,miotto2024zepyros}.\\
More recently, we extended the same formalism to another property that can be described with numerical values assigned to each surface point: the electrostatic potential \cite{Grassmann2023}. We demonstrated that interfaces also exhibit electrostatic complementarity.
To evaluate it, we applied the Zernike method to molecular surfaces for which the electrostatic potential had been calculated through the Poisson-Boltzmann equation.\\
Another fundamental contribution to protein binding is the degree of hydrophilicity and hydrophobicity at the interfaces: interacting sites must present a balance between the former, ensuring the structural stability of each partner, and the latter, which is need to ensure binding.
In previous works, we developed a hydropathy scale that assigns a hydropathy index $H$ to each amino acid based on changes observed in the hydrogen bond network of water molecules surrounding each given compound during Molecular Dynamics (MD) simulations \cite{DiRienzo2021}.\\

Here, we present a novel method to evaluate shape, electrostatic, and hydropathy complementarity through a unified computational protocol. By characterizing the radial variability of these aspects at interfaces, we find that they are maximized within an area of 10\AA~radius. We define new binding matrices that compactly describe shape, electrostatic, and hydropathy complementarity between surface patches of this size. These matrices are used to train a Neural Network (NN) to identify interacting residues at the core of binding sites.\\
The proposed Core Interacting Residues Network (CIRNet), tested on a balanced dataset of 1086 complexes, achieves an accuracy of 0.82 in identifying core interacting residue pairs. In a blind-search on a second dataset of 100 complexes, CIRNet distinguishes core interacting residues from all other pairs with a ROC AUC of 0.72.\\
To evaluate CIRNet's performance, we used its predictions as additional information for selecting the most reliable models from three publicly available docking servers: ClusPro \cite{comeau2004cluspro}, PyDock \cite{cheng2007pydock}, and LzerD \cite{christoffer2021lzerd}.\\
ClusPro was introduced in 2004 and in the following years has been constantly improved \cite{comeau2007cluspro, kozakov2010achieving, kozakov2013good, kozakov2017cluspro}. It performs a rigid-body docking and then clusters the multiple docked conformations to identify the most probable complex structures. To enhance the accuracy of its predictions, it integrates various scoring functions, including van der Waals, electrostatic, and desolvation energies. Although the most near-native structure is not always included in the largest cluster, it has been shown that the 30 largest clusters include near-native structures for 92\% of complexes in a protein–protein benchmark set \cite{comeau2004cluspro}. Another rigid-body docking is performed by pyDock, whose energy-based scoring system is focused on Coulombic electrostatics with distance dependent dielectric constant and implicit desolvation energy. 
Its scoring was able to identify a near-native solution within the top 20 or 100 solutions in 37\% and 56\% of the cases respectively \cite{cheng2007pydock}.
LzerD, on the other hand, is a flexible protein-protein docking server that utilizes geometric hashing and 3D Zernike descriptors to predict complex structures. For bound docking predictions, LzerD finds a near-native solution within the top 1000 ranks in 74\% of cases \cite{venkatraman2009protein}.\\
When tested on 30 complexes, we observed that the docking scores did not always reflect the similarity to the native complex. 
Using the CIRNet prediction as an additional filter for the docking poses, we were able to extract some of the worst poses among the first ten proposed by ClusPro, PyDock, and LzerD.\\
Our results show that the NN predictions for core interacting pairs of poses that strongly differ from the native structure are usually lower. Thus, by removing poses whose core interacting residues are not recognized as such by CIRNet, we can partially filter out structures that despite being highly ranked are far from the native pose. Even when the true conformation does not present core interacting residues, CIRNet can still identify pairs that are proposed as core interacting residues by the docking but do not meet the physical and geometrical criteria for core binding regions.

\section*{Results}

\subsection*{Composition of protein-protein interfaces}
\begin{figure*}[ht]
\centering
\includegraphics[width=\linewidth]{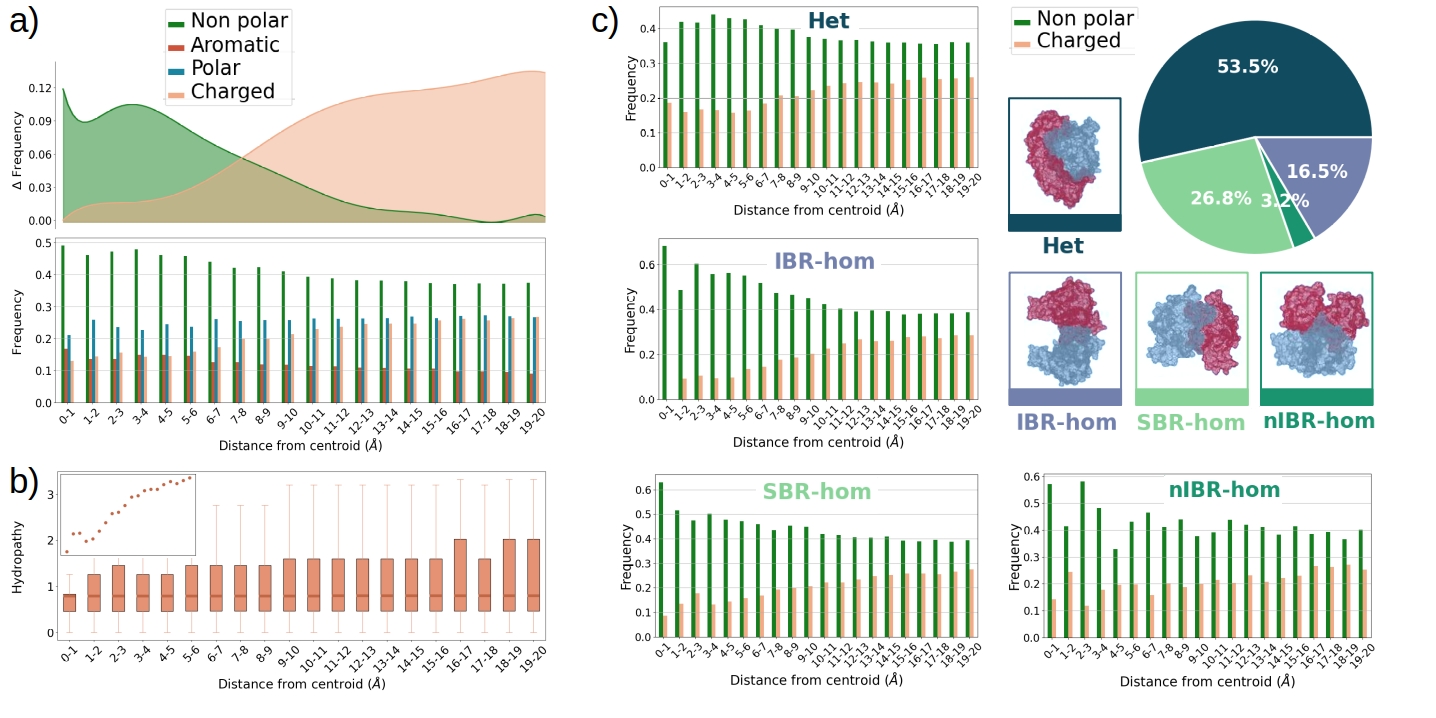}
\caption{\textbf{Amino acid composition, hydropathy properties, and dimer classification of the dataset interfaces.}
\textbf{a)} On bottom the frequency of each of the four classes of amino acids (non-polar, aromatic, polar, and charged, represented in green, red, blue, and yellow respectively) in 1 \AA-wide annular regions at increasingly larger distances from the center of the binding sites. For non-polar and aromatic residues, the difference between the frequency value in each annular region and the minimum frequency of that class is shown on top.
\textbf{b)} The box-plot displays the $H$ values of the residues included in the annular regions defined in a). The inset shows the mean $H$ value for each annular region.
\textbf{c)} The dataset includes heterodimers and homodimers with identical, non-identical, and shifted binding regions, as shown by the percentages in the pie plot. An example for each class is provided in the four inserts. The frequencies of non-polar and charged residues shown in a) are divided according to these classes.
}
\label{fig1}
\end{figure*}

To perform a quantitative wide-ranging spatial characterization of the interfaces that aimed to predict, we used a balanced dataset already used by Milanetti  \textit{et al.} \cite{Milanetti2021} counting 3721 complexes with available structural data. 
The dataset includes 1730 homodimers, which can be categorized into 614 dimers with an Identical Binding Region (IBR-hom), 998 Shifted Binding Region (SBR-hom), and 118 non-Identical Binding Region (nIBR-hom), based on the similarity of the interacting patches. Additionally, the dataset contains 1991 heterodimers (nIBR-het).\\ 
For each complex, we evaluated the chemical features of the surface regions surrounding the geometrical center of the interfaces, divided into annular regions with a 1\AA~radius. 
We defined the interfaces as the "hot spots" on a protein surface that are within 6\AA~of its partner surface.
Given that the interfaces in our dataset have an average maximum radius of 17 \AA, we analyzed the contributions of amino acids within a 20\AA~radius from the center of the interfaces.\\
As shown in Figure \ref{fig1}a), the most frequent solvent-exposed amino acids are hydrophobic. Hydrophobic residues are particularly predominant at the core of the binding regions compared to the other classes.
Up to a distance of 10\AA~from the interface centroid, they oscillate between approximately 50\% and 40\% of the total. We define this region as the core of the binding site.
At the rim of the binding sites, hydrophobic residues become less abundant, although they remain the predominant class with a frequency of about 37\%. 
The opposite trend is observed for charged residues: at the core of the binding site, they account for less than 20\% of the total residues, while at the rim, their frequency increases by one-fourth.
Electrostatic interactions seem to guide distant proteins as they approach each other and reorient to find their binding sites. Once positioned, core residues lead the final interlocking. These residues can establish strong hydrophobic interactions when they have low hydropathy ($H$) values, as defined by the scale proposed by Di Rienzo \textit{et al.} \cite{DiRienzo2021}. Figure \ref{fig1}b) shows that core residues indeed tend to have smaller $H$ values.\\
Figure \ref{fig1}c) shows that the core and rim regions of the interfaces have similar trends across all four dimer classes. In all cases, non-polar residues are particularly predominant at the core, while the frequency of charged amino acids increases outside a 10\AA~radius. Non-polar residues are more frequent at the interfaces of homodimers than heterodimers; for homodimers, shape complementarity seems to be a less stringent requirement for binding.

\subsection*{Complementarity at protein-protein interfaces}
\begin{figure*}[ht]
\centering
\includegraphics[width=\linewidth]{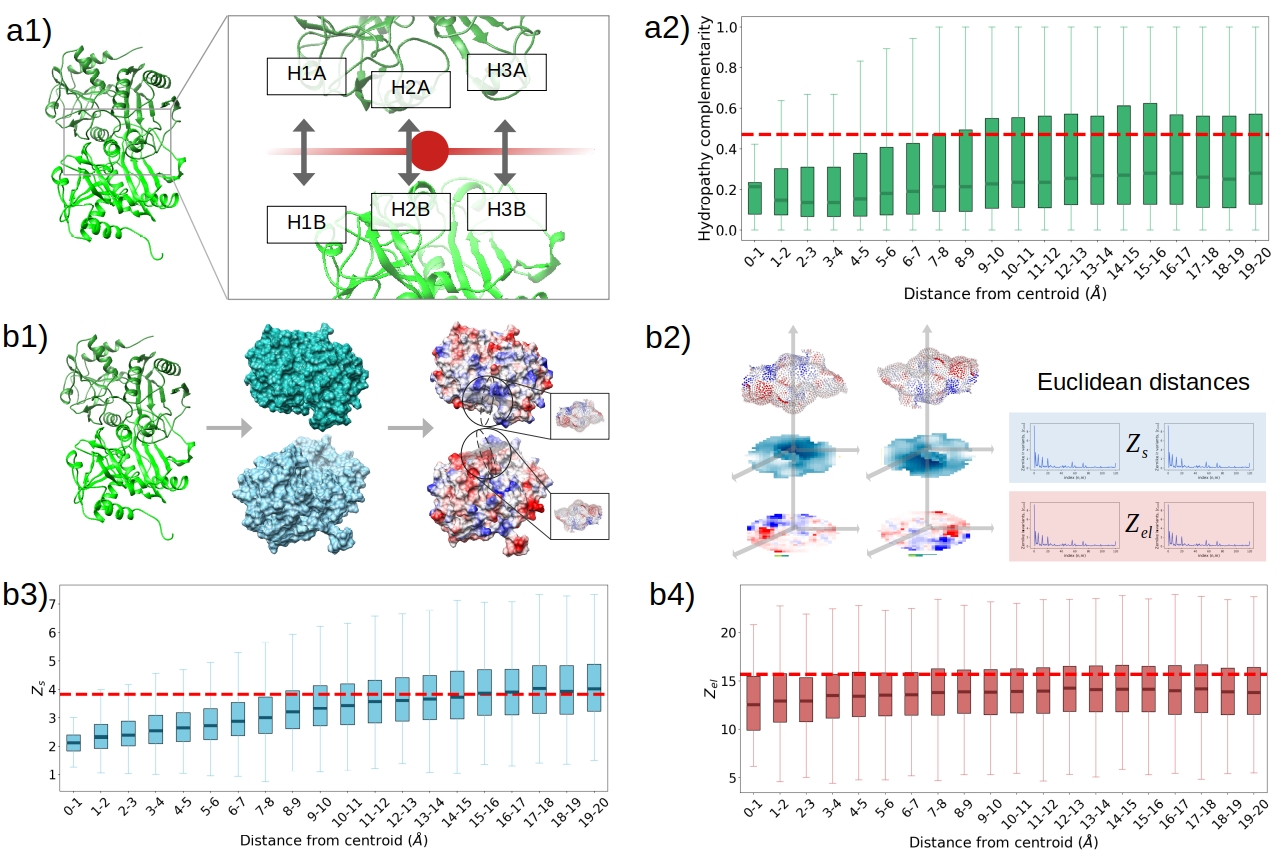}
\caption{\textbf{Characterization of the hydropathy, shape, and electrostatic complementarity at the interfaces.}
\textbf{a1)} The center of each binding site is identified. Starting from the $H$ index of the interacting residues facing each other (closer than 3\AA) the $Hr$ for each residue pair is computed. 
\textbf{a2)} Box-plot of the $Hr$ values of interacting residue pairs in 1\AA-wide annular regions at increasingly larger distances from the center of the interfaces. The dotted red line indicates the mean $Hr$ value obtained between random non-interacting pairs.
\textbf{b1)} Molecular representations of two proteins forming a dimer.  The solvent-exposed and electrostatic potential surfaces are obtained from the secondary structure. In the latter, each point is colored according to its electrostatic potential value.
Spheres with a 9\AA~radius are used to select a patch on the surface: in the zoomed portion, the selected points.
\textbf{b2)} 2D projections of the patches selected in b1). In the blue scale, the shape projection: the color of each pixel is determined by the distance $r$ from a predefined origin (see Methods for details) of the surface points projected inside of it.
In the blue-red scale, the electrostatic projection: colors are determined by the electrostatic potential values of the projected points. 
Each projection is a 2D function that can be associated with a Zernike vector. The Euclidean distances $Z_s$ and $Z_{el}$ between these vectors quantify the complementarity, in terms of shape or electrostatic, of the two corresponding patches.
\textbf{b3)} Box-plot of the $Z_s$ values of interacting patches in 1\AA-wide annular regions at increasingly larger distances from the center of the interfaces. The dotted red line indicates the mean $Z_s$ value obtained between random non-interacting patches.
\textbf{b4)} Same as in b3), but for $Z_{el}$.}
\label{fig2}
\end{figure*}

To further characterize the hydrophobic interactions at the interfaces, we compared the $H$ value of residues facing each other and evaluated the strength of these interactions at increasing distances from the center of the interfaces (see Figure \ref{fig2}a1)).\\
The bond between amino acids depends on their hydrophobic or hydrophilic nature. Interactions between hydrophobic residues (low $H$ values) are strong because such bonds are energetically favorable, reducing the exposure of hydrophobic surfaces to the aqueous environment. Strong interactions occur between hydrophilic residues (high $H$ values) as well, since they can form hydrogen bonds. However, when a hydrophobic residue interacts with a hydrophilic residue, there is no tendency for a strong interaction.\\
Given these interaction tendencies, we can model the bond strength between two residues with their $H$ indices. To quantify this strength, we defined the hydropathy complementarity between two residues $A$ and $B$ as:
\begin{equation}
    Hr = -4 \times \bigg(\frac{H_A H_B}{11.0224}\bigg)^2+\bigg(\frac{H_A H_B}{11.0224}\bigg),
\end{equation}
where $H_A$ and $H_B$ are the hydropathy indices of the two amino acids.
$Hr$ reflects the idea that residues with similar hydropathy values (both high or low) have stronger interactions compared to those between residues with opposing hydropathy characteristics.
Values close to zero correspond to strong interactions: the closer $Hr$ is to one, the weaker the bound between the residues.
Interface residues tend to be hydrophobic: the mean hydropathy complementarity between random non-interacting residues is $Hr=0.47$, as shown by the red dotted line of Figure \ref{fig2}a2). Close to the interface centroid (residues closer than 4\AA), $Hr$ has the lowest value. The rest of the core region shows a slow increase of $Hr$, which then reaches and maintains a stable value at the rim.
Since protein binding depends on an interplay between various contributions on the molecular surface, we extended this analysis to two more features: shape and electrostatic complementarities.
For the former, it is widely known that the side-chain optimization of the short-ranged van der Waals interactions between interfaces leads to a local shape complementarity of the proteins’ molecular surfaces. 
The role of electrostatic interactions, including hydrogen bonding, ionic/Coulombic, cation$-$ $\pi$, $\pi-\pi$, lone-pair sigma hole, and orthogonal multipolar interactions~\cite{bauer2019electrostatic, Sheinerman2002, DiRienzo2021},
is more debated \cite{Sheinerman2002,bauer2019electrostatic}: while it is understood that they can bring close distant partners and reorient them \cite{Shashikala2019}, the role they play in binding on small distances is still uncertain \cite{zhang2011role,vascon2020protein,kundrotas2006electrostatic,zhou2018electrostatic,yoshida2019exploring}.
Recently, we characterized the role of these interactions in protein binding in surface regions with a radius of 9\AA, where shape complementarity is maximized \cite{Grassmann2023}. We proposed a computational protocol that can distinguish between interacting and non-interacting regions by describing their electrostatic potential surfaces with vectors and comparing these descriptors. 
This method derives from the 2D Zernike method, proposed in 2021 to quickly evaluate the shape complementarity at interfaces \cite{Milanetti2021}. In a blind search, the Zernike protocol can identify binding regions in 60\% of cases. 
When applied to the electrostatic potential surface, the protocol can discriminate between transient and permanent interactions with a ROC AUC of 0.8.\\
The main steps of the Zernike-based method are shown in Figure \ref{fig2}b1),b2). 
Starting from a protein structure, we compute the solvent-exposed surface and the corresponding electrostatic potential surface, where each surface point is described by its spatial coordinates and by the value of the electrostatic potential generated by the protein in that point. See the Methods Section for more details.
Next, we define a surface patch by selecting all the points on the electrostatic potential surface contained in a sphere with a 9\AA~radius centered on a surface point.
This patch is then projected on a plane so that each patch is associated with two 2D matrices. In the first one, each pixel derives from the spatial positions of the points projected inside it. In the second matrix, each pixel is the mean of the electrostatic potential values of the projected points. 
These matrices are 2D functions that can be decomposed in the basis of the Zernike polynomials.
The Euclidean distance between the Zernike vectors
describing two patches quantifies their complementarity, in terms of shape or electrostatic.\\
Figure \ref{fig2}b3) depicts the Euclidean distances $Z_s$ between the Zernike vectors describing the shape of the surface patches relative to residues at increasing distances from the interface centroid. Lower $Z_s$ values indicate higher shape complementarity (i.e., higher binding propensity \cite{Milanetti2021}).
The core region is clearly distinguished from the rest of the interface by higher shape complementarity values compared to the mean $Z_s$ between random non-interacting residues (red dotted line).
This distinction is less marked for electrostatic complementarity, as shown in Figure \ref{fig2}b4). 
The level of electrostatic complementarity seems to be uniform between core and rim.\\
However, both regions have lower Euclidean distances $Z_{el}$ between the Zernike vectors describing the electrostatic potential of the surface patches compared to that between random non-interacting residues.\\

These observations allow us to quantitatively define the typical binding site as the region where shape complementarity and hydrophobic contributions are maximized. This region has a radius of approximately 10\AA.

\subsection*{Machine learning for residues pairs classification}
\begin{figure*}[ht]
\centering
\includegraphics[width=\linewidth]{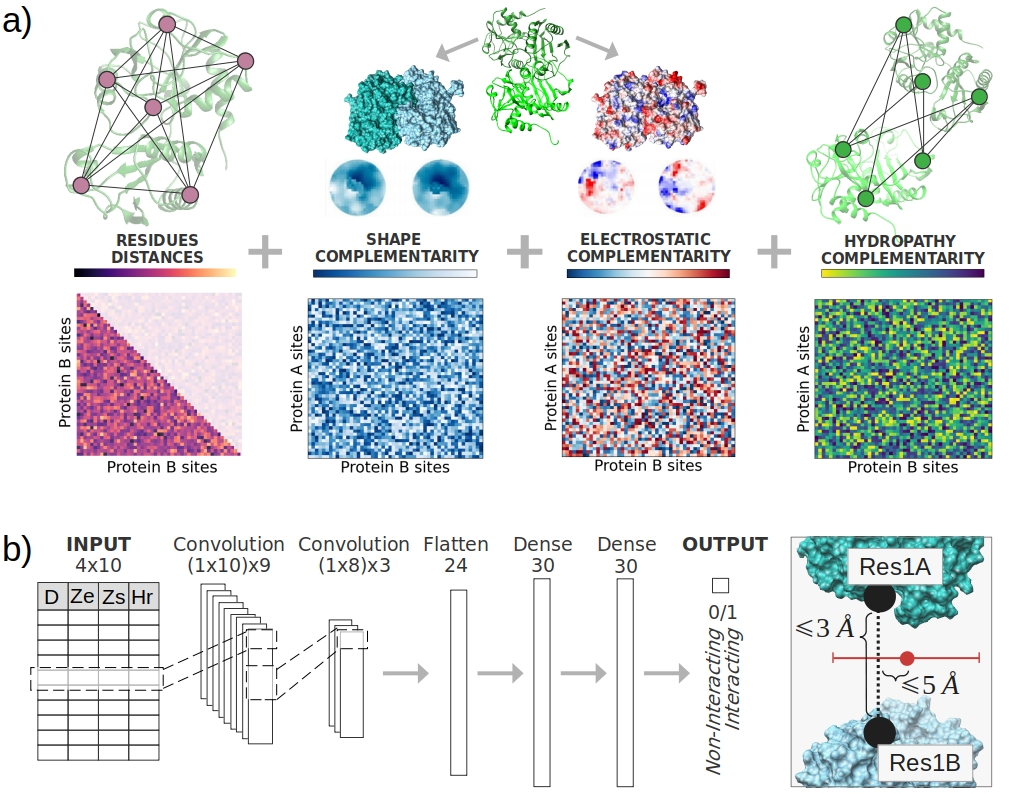}
\caption{\textbf{Architecture of CIRNet and input data.}
\textbf{a)} $Z_s$, $Z_{el}$, and $Hr$ are computed for each possible pair of residues. 
This process results in three matrices for each protein pair, representing shape, electrostatic, and hydropathy complementarity between all residues (blue, red-blue, and blue-green color scales). For one of the proteins a matrix with the distances between all of its residues (pink color scale) is defined as well.
\textbf{b)} Each residue pair is associated with a $4\times10$ matrix that summarizes the shape, electrostatic, and hydropathy complementarity levels between the first analyzed residue and the neighborhood of the second one. This matrix is fed into a NN with two convolutional layers and two dense layers. The NN then classifies each residue pair based on its probability of being at the core of an interface. Core interacting residues are defined as those closer than 3\AA~and within 5\AA~from the interface center.}
\label{fig3}
\end{figure*}

We have discussed how the essential interactions for binding stability are maximized in a core region at the interface with a radius of approximately 10\AA.
We have also shown that the Zernike method can characterize each region on the protein surface in terms of shape and electrostatic potential, enabling rapid comparison between different patches. 
Leveraging these results, we introduce a NN architecture that can identify pairs of core interacting residues by combining the characterization of shape, electrostatics, and hydropathy complementarities.\\
Figure \ref{fig3}a) illustrates the procedure leading to the data given as input to CIRNet.
Given a pair of proteins, we compute for each possible residues pair $Z_s$, $Z_{el}$, and $Hr$. This results in three binding matrices with sizes $N\times M$, where $N$ and $M$ are the number of residues in the first and second protein, respectively. For the second protein (protein $B$), we also compute the distance between all of its residues.
This fourth matrix identifies the first nine neighbors of each residue in $B$. This number of neighbors defines an area with an average radius of 10\AA, corresponding to the core regions expected to be binding discriminants.
For each pair of residues 1$A$, 1$B$, we build a matrix with four columns and ten rows. The first row has the shape, electrostatic, and hydropathy complementarity value between 1$A$ and 1$B$. The fourth value is the distance of residues 1$B$ from itself (0). The following rows have the complementarities between residue 1$A$ and the closest neighbor of 1$B$, 2$B$, as well as the distance between 2$B$ and 1$B$. The second row refers to the second neighbor, and so on.\\
The resulting matrix is given as input to CIRNet, whose architecture is shown in Figure \ref{fig3}b) and described more in-depth in the Methods Section. CIRNet assigns to each pair of residues a value reflecting the probability of it being a pair of interacting residues at the core of the interfaces. “True” labels are assigned to the residues that are closer than 3\AA~and that have a distance $R$ from the centroid of the interface smaller than 5\AA. These parameters resulted in the best training of the network.
Table \ref{table1} shows the ROC AUC between the distributions of the classification of core interacting pairs and all the others for increasing $R$ values. Smaller $R$ were not considered to avoid having too few cases of core interacting pairs.

\begin{table}[htbp]
\centering
\begin{tabular}{|c|c|}
\hline
\textbf{R} & \textbf{ROC AUC} \\
\hline
$<$5\AA & 0.78 \\
$<$6\AA & 0.78 \\
$<$7\AA & 0.74 \\
$<$8\AA & 0.76 \\
\hline
\end{tabular}
\hspace{2em} 
\begin{tabular}{|c|c|}
\hline
\textbf{R} & \textbf{ROC AUC} \\
\hline
$<$9\AA & 0.70 \\
$<$10\AA & 0.71 \\
$<$11\AA & 0.70 \\
$<$12\AA & 0.68 \\
\hline
\end{tabular}
\caption{\textbf{NN classification efficacy depending on data labeling.}
ROC AUC of the NN classification between real interacting and non-interacting residue pairs for increasing maximum distances $R$ from the interface center, within which residue pairs are labeled as interacting.}
\label{table1}
\end{table}

\begin{figure*}[ht]
\centering
\includegraphics[width=\linewidth]{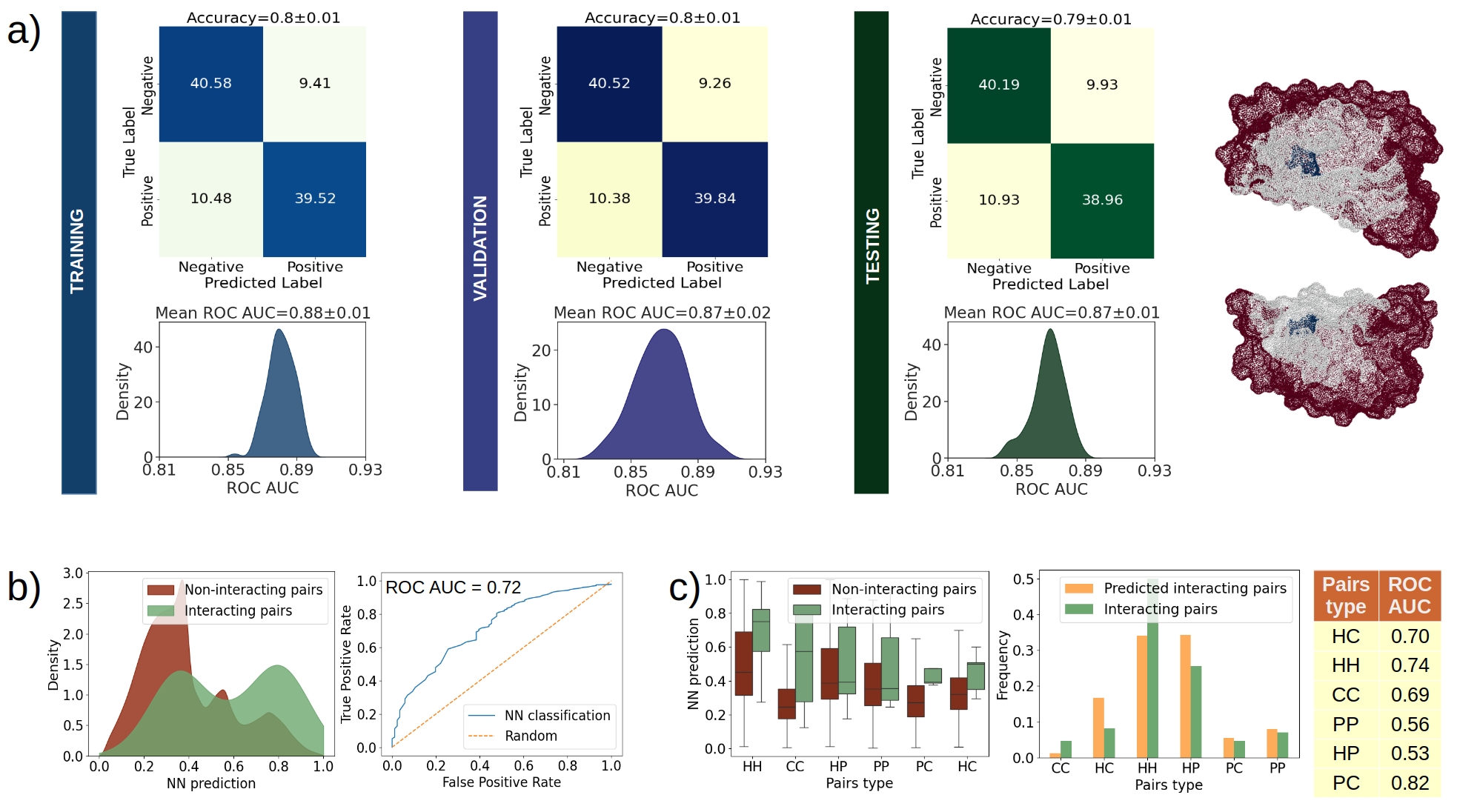}
\caption{\textbf{Accuracy of CIRNet during training and testing, and blind-search  performance for core interacting residues based on chemical significance.} 
\textbf{a)} The mean results for 100 repetitions of CIRNet training, validation, and testing are shown in blue, violet and green respectively. For each phase, the mean confusion matrix is presented along with the average accuracy. The distributions of ROC AUC values for each repetition are displayed, reflecting the classification performance between core interacting residues and decoys. The mean ROC AUC for these distributions is indicated above each plot. An example complex is provided on the right, showing two interacting residues (in blue) at the core of an interface (in grey) identified by the network as the most interacting.
\textbf{b)} On the left, the CIRNet predictions from a blind-search for core interacting residues across a dataset of 100 complexes are displayed. The NN classifications are categorized into true core interacting residues (green) and decoys (red). On the right, the ROC curve between these two distributions is shown, with the AUC value reported.
\textbf{c)} On the left, the distributions from b) are further stratified based on the chemical nature of the residues. Pairs are organized according to the mean NN prediction value for core interacting residues. The barplot on the right illustrates the frequency of each residue class among both true core interacting residues and those identified as such by CIRNet. Additionally, ROC AUC values for the distributions stratified by residue type are provided.}
\label{fig4}
\end{figure*}

CIRNet was trained on 2535 complexes, with 20\% of this dataset used for validation at each epoch. To determine the optimal NN parameters, we created a balanced benchmark with an equal number of core interacting residues and negative examples. The network was subsequently tested on a balanced set of residue pairs from an additional 1086 complexes.
Figure \ref{fig4}a) displays the average results from 100 repetitions of the training, validation, and testing procedures. In each repetition, complexes for the three phases were randomly selected from the full dataset of 3721 complexes described in the Methods Section. CIRNet achieved a mean accuracy of approximately 0.8 during training and validation, with a higher performance in identifying non-core interacting residues compared to core interacting pairs. This is evidenced by a higher rate of True Negatives relative to True Positives. Despite this, CIRNet distinguished between the two classes with a mean ROC AUC of about 0.87. The performance metrics were consistent across the 100 test repetitions, with a mean accuracy of around 0.79 and an ROC AUC of approximately 0.87.\\
Subsequently, we assessed CIRNet's ability to blind-test core interacting residues on 100 complexes that were excluded from the training dataset and had not been presented to the NN before. In this evaluation, all possible residue pairs were classified by CIRNet, yielding the distribution shown in Figure \ref{fig4}b). Core interacting residues were distinguished from other pairs with a ROC AUC of 0.72. The NN prediction threshold was optimized to maximize the True Positive Rate while minimizing the False Positive Rate: only pairs with a prediction score above 0.62 were classified as core interacting residues.\\
To further analyze the double-peaked distribution of core interacting residues in the NN prediction, depicted in Figure \ref{fig4}b), we investigated the chemical significance of the network's learning. Specifically, we stratified the classification of pairs based on residue types (hydrophobic, charged, and polar).
The boxplot in Figure \ref{fig4}c) reveals that the left peak in the distribution corresponds to hydrophobic-charged (HC) and polar-charged (PC) pairs, with prediction values centered around 0.55 and 0.4, respectively. The right peak is primarily due to hydrophobic-hydrophobic (HH) pairs, with a mode around 0.8. HH pairs are the most abundant among the true-labeled ones, constituting approximately 50\% of the true-labeled core interacting residues. Although the network fails to correctly classify about one-third of these pairs (as shown by the barplot in Figure \ref{fig4}c)), the ROC AUC for HH pairs, when distinguishing core interacting residues from negative examples, is among the highest, reaching 0.74, similar to the 0.82 achieved for PC pairs. In contrast, polar-polar (PP) and hydrophobic-polar (HP) pairs are less distinctly classified as core interacting residues or decoys, as indicated by ROC AUC values of 0.56 and 0.53, respectively.

\subsection*{Docking poses selection}
\begin{figure*}[ht]
\centering
\includegraphics[width=\linewidth]{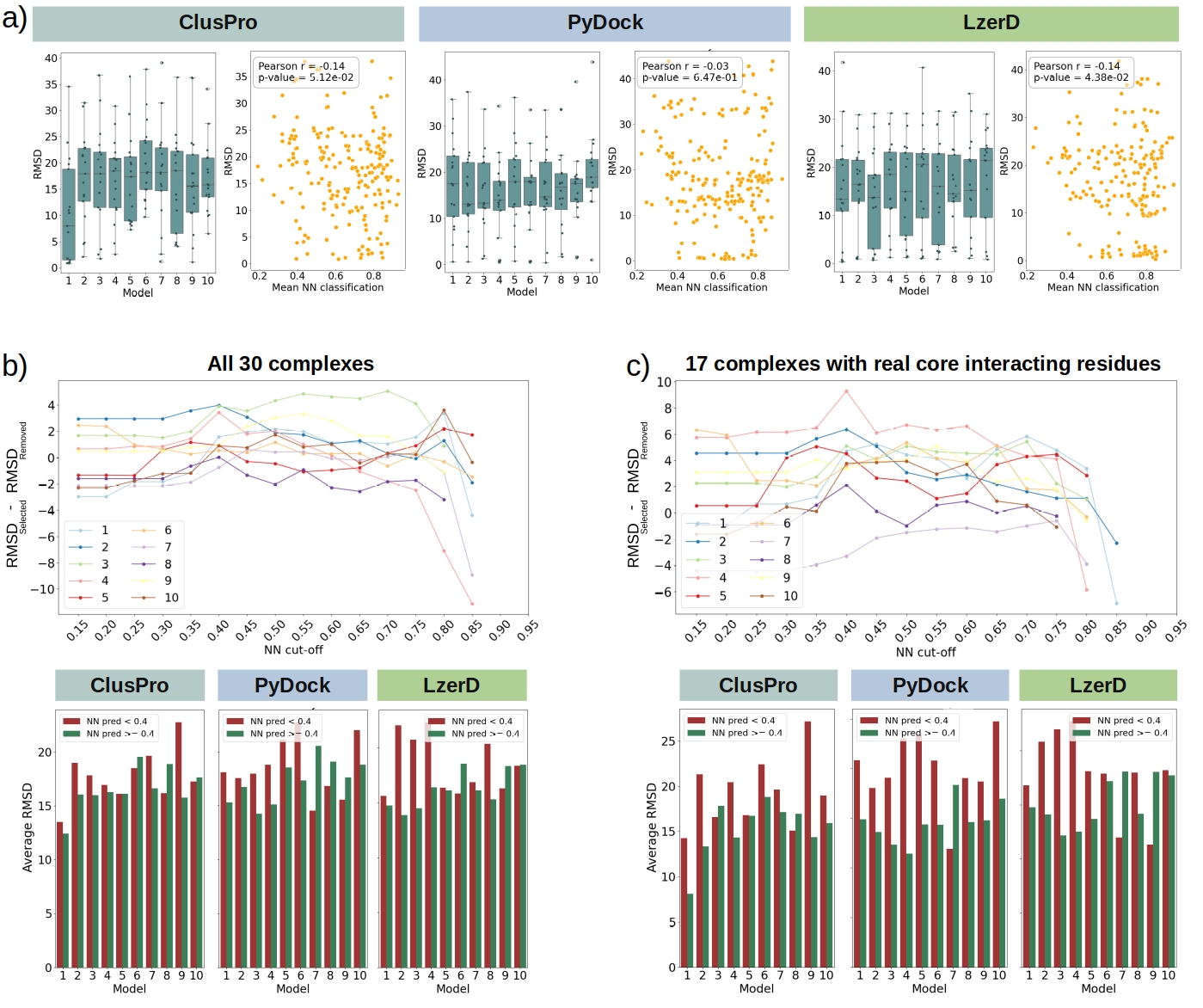}
\caption{\textbf{Ranking performance on the first ten models of the ClusPro, PyDock, and LzerD docking servers and NN classification of their core interacting pairs used as a re-ranking score.} 
\textbf{a)} The individual units of 30 complexes with known bind structures were docked using three publicly available tools: ClusPro, PyDock, and LzerD (from left to right). For each docking server, the RMSD between the first ten models and the native complex is computed. Each subplot reports the RMSD of the docked structures as a function of the docking ranking (left) and as a function of the NN classification of core interacting residues at the interface (right). Pearson correlations and p-values are also reported for the latter.
\textbf{b)} For all docked poses, the residue pairs within 3\AA~and 5\AA~from the interface centroid are selected. The NN prediction scores is calculated for these pairs and various NN prediction thresholds are applied to classify pairs as core interacting. The top plot shows the difference in RMSD between poses associated with residue pairs classified as core interacting (RMSD$_{Selected}$) and those classified as non-core interacting (RMSD$_{Removed}$) for increasing NN prediction thresholds. 
The RMSD is averaged for each of the ten ranks, indicated by the legend.
The bottom plot shows the average RMSD of poses selected (green) or removed (red) using an NN prediction threshold of 0.4, averaged across all poses in each ranking level for each docking server.
\textbf{c)} Same as in b), but only the complexes with core interacting residues in their native pose are considered.
}
\label{fig5}
\end{figure*}

Having verified CIRNet’s effectiveness in identifying core interacting residues through blind searches, we next evaluated its efficiency as a re-ranking tool for docking results. Docking techniques face the challenge of efficiently ranking hundreds-of-thousands to millions of possible configurations to identify those closest to the native complex. The need for rapid computation often compromises chemical accuracy, leading to known errors in docking scoring functions \cite{irwin2016docking,mobley2009binding,bender2021practical}.\\
We tested CIRNet on three publicly available docking servers: ClusPro, PyDock, and LzerD. ClusPro and PyDock are rigid-body docking methods, while LzerD performs flexible docking. For a dataset of 30 complexes with known structures, we docked their individual units using each server and selected the top ten models from each. We then computed the RMSD between each model and the native pose and identified residues on each partner within 3\AA~and 5\AA~from the interface centroid, corresponding to our definition of core interacting residues. CIRNet provided interaction prediction scores for these selected pairs.\\
Figure \ref{fig5}a) illustrates that docking ranks do not consistently reflect similarity to the native pose. For ClusPro, the median RMSD of the top-ranked pose (model 1) is approximately 8\AA, while other ranks show medians between 15 and 18\AA. PyDock’s lowest median RMSD ($\sim$13\AA) occurs in the second, third, and fourth models, with other ranks ranging between 16 and 18\AA. LzerD performs the worst, with medians between 13 and 21\AA, not clearly identifying the first models as the best ones.\\
The average CIRNet prediction score for core interacting pairs does not show a strong correlation with RMSD (see again Figure \ref{fig5}a)). However, higher RMSD poses tend to have lower NN prediction scores, while lower RMSD poses generally receive higher scores, as indicated by a Pearson correlation of -0.14 for ClusPro and LzerD, and -0.03 for PyDock.\\

To assess if CIRNet can improve docking pose selection, we analyzed the impact of removing poses not recognized as core interacting residues by CIRNet.\\
Figure \ref{fig5}b) shows the difference in RMSD between poses associated with core interacting residues (RMSD$_{Selected}$) and those removed (RMSD$_{Removed}$) for various NN prediction thresholds. A threshold of 0.4 proves most effective at removing poorer poses from the first, second, and fourth ranks, with a notable improvement of nearly 4\AA~for the third rank as well.\\
The bottom plot in Figure \ref{fig5}b) details the average RMSD of removed (red) versus selected (green) poses according to this threshold: the former tend to have a higher average RMSD.
For ClusPro, 40-50\% of poses are removed, with significant RMSD reductions for ranks two, three, seven, and nine (8\%, 7\%, 12\%, and 25\%, respectively). PyDock shows similar removal rates (20-50\%), with RMSD reductions of about 10\%, 15\%, 19\%, 10\%, and 23\% for ranks one and three through six. LzerD removes fewer poses, conserving around 70\% of structures in the top four ranks, but achieves the most significant RMSD improvements (20-30\%) for the second, third, and fourth models.\\
Of the 30 complexes, 13 lack core interacting residues in their native poses. Figure \ref{fig5}c) shows that when only the other 17 complexes are considered, the RMSD difference between poses removed and selected by CIRNet is greater for most ranks. The optimal NN prediction threshold remains 0.4. Using this threshold, similar percentages of poses are removed as before, but with substantially reduced average RMSD. ClusPro’s first two ranks see reductions of approximately 39\% and 25\%, respectively. PyDock’s reductions are between 15\% and 23\% for the top three ranks, with a 62\% reduction for the fourth rank. LzerD shows reductions of about 10\%, 30\%, 58\%, and 36\% for ranks one through four, respectively.\\

We can conclude that CIRNet demonstrates optimal performance for complexes with close-range core interacting residues but remains a valuable tool for improving docking pose selection across different types of dimers by effectively removing the least accurate predictions.

\section*{Discussion}
Understanding the molecular mechanisms underlying protein binding is a fundamental pursuit in molecular biology, with implications for protein design and the mapping of the human interactome \cite{zhang2012structure, hermann2007structure, kortemme2004computational, Miotto2024zolfo}. A critical aspect of this endeavor is the accurate characterization and prediction of interacting regions.\\
The properties of the environment, such as pH and ionic strength, as well as the presence of ordered water molecules, and the complexity of protein–protein interactions make the complete characterization of binding sites a demanding problem. Docking servers are instrumental in predicting the most likely poses of protein complexes, yet their performance is often limited by scoring functions that lack sufficient accuracy. Consequently, these scores frequently do not reflect how closely the proposed complex approximates the native pose.\\
To address this challenge, we propose a computational strategy that leverages a unified formalism based on the 2D Zernike orthogonal polynomial basis to describe the molecular and electrostatic potential surfaces of proteins. Lennard-Jones and Coulomb potentials, which represent short-range van der Waals forces and long-range electrostatic forces respectively, play crucial roles in protein interactions. These potentials are vital for both the initial recognition of binding partners and subsequent binding adaptation.\\
Our analysis reveals that interfaces exhibit a core region with a radius of approximately 10\AA~where shape complementarity is maximized. This core region also shows higher hydrophobicity complementarity compared to the rest of the surface.\\
We combine information on these three aspects with a shape descriptor to train CIRNet, a neural network designed to identify pairs of interacting residues situated at these core interfaces. In a blind search, CIRNet distinguishes core interacting pairs (residues within 3\AA~ and 5\AA~from the interface centroid) from negative examples, achieving a ROC AUC of 0.72.\\
Furthermore, we tested CIRNet as a re-ranking method for poses proposed by three publicly available docking servers: ClusPro, PyDock, and LzerD. Our results show that the average NN predictions for core interacting pairs tend to be lower (indicating a lower probability of interaction) for poses that are far from the native structure (high RMSD). By removing poses where core interacting residues are not recognized as such by CIRNet, we were able to partially filter out structures that, despite high ranking, are far from the native pose. Although the best results were achieved for complexes where the native state features closely interacting core residues, CIRNet’s protocol can be extended to other types of complexes. Even when true binding does not rely solely on core interacting residues, CIRNet can still identify pairs that do not meet the physical and geometrical criteria for core binding regions.

\section*{Methods}

\subsection*{Protein complex dataset}
To probe the physical and chemical features of protein-protein binding regions, we collected a dataset of protein-protein dimers with structural information obtained through X-ray crystallography, sourced from the 3D Complex Database \cite{Levy_2006}.
Starting from the dataset already used by Milanetti \textit{et al.} \cite{Milanetti2021}, we selected only pair interactions with known experimental pH values,  resulting in a dataset of 3721 complexes.
These dimers can be classified into four groups based on the chains composition and spatial orientation:
\begin{itemize}
    \item 614 homodimers with Identical Binding Regions (IBR-hom): binding regions with at least 70\% residues overlap.
    \item 998 homodimers with Shifted Binding Regions (SBR-hom): interacting patches with 30\% to 70\% of common residues.
    \item 118 homodimers with non-Identical Binding Regions (nIBR-hom): binding regions sharing less than 30\% of residues.
    \item 1991 heterodimers (nIBR-het): complexes involving interactions between two different proteins.
\end{itemize}

We then removed 100 complexes from this dataset, leaving 3621 for identifying the optimal parameters of a NN to pinpoint core interacting residues and to test its performance on a balanced dataset. The resulting architecture was subsequently applied in a blind search over the 100 removed complexes. Of these, 30 complexes were randomly selected and docked using three different docking software: ClusPro, PyDock, and LzerD.\\

It is noteworthy that approximately 60\% of the complexes exhibit core interacting residues at their interfaces. We included complexes lacking this characteristic in the dataset to test CIRNet's applicability across different types of complexes. This allowed us to assess CIRNet’s performance in identifying core residues as well as to use these non-core complexes as negative examples or to evaluate CIRNet’s ability to exclude residue pairs that cannot be at the core of the interfaces.

\subsection*{Computation of the surfaces and binding sites definition}
For each protein in the dataset, the solvent-accessible surface was computed using DMS \cite{Richards1977AreasVP}, with a resolution of 5 points per \AA$^2$ and a water probe radius of 1.4 \AA. 
The electrostatic potential was calculated with the APBS code \cite{Jurrus2017}, taking into account the experimental pH and treating each protein independently from its partner.
To define the electrostatic potential surface, we created a grid and assigned electrostatic potential values to each grid cell corresponding to surface points.\\
Binding sites were defined as regions on a protein surface that are within 6 \AA~of the partner surface. The center of an interface was determined as the centroid of the surface points within this binding region.

\subsection*{Hydropathy scale}
Di Rienzo \textit{et al.} \cite{DiRienzo2021} introduced a method to compactly represent and compare the hydrophobicity of molecular surface patches. They conducted MD simulations for each of the 20 natural amino acids and analyzed the variations in the hydrogen bond network of surrounding water molecules. Observing the spatial reorganization in the local structure allowed them to evaluate their hydrophilicity and hydrophobicity features and associate each amino acid with a hydropathy index (H). The resulting hydropathy scale considers the collective response of protein hydration waters to the local nanoscale chemical and topographical patterns.


\subsection*{Patch definition and projection}
A surface patch is defined by placing a sphere with a radius of $R=$9\AA~centered on a surface point and selecting the surface points within this sphere. Previous studies \cite{Milanetti2021, Grassmann2023} identified that this range of $R$ values is optimal for accurately identifying binding regions based on shape and electrostatic complementarity.\\
Next, we fit a plane to the selected patch points and re-orient the patch so that its normal vector is perpendicular to the plane. To compare shape complementarity between two patches, they must be re-oriented such that one patch has its solvent-exposed side facing the positive z-axis and the other facing the negative z-axis.\\
Each point on the re-oriented patch is described by its three spatial coordinates and the electrostatic potential at that point. These descriptions are then projected onto the x-y plane.\\
We define a point $C$ on the z-axis and set it such that the angle $\theta$ between the z-axis and any secant connecting $C$ to a point on the patch is $\theta=45^\circ$. To obtain the shape projection, each surface point is labeled with its distance $r$ from $C$. A 25$\times$25  pixels grid is constructed, and each pixel is associated with the mean $r$ value calculated on the points projected inside it.\\
For the electrostatic projection, a separate 25$\times$25 grid is created, and each pixel is assigned the mean electrostatic potential of the points projected onto it.\\
The parameters used for these projections were identified as the most effective in the previously mentioned work \cite{Milanetti2021}.

\subsection*{Zernike 2D protocol}
Any function of two variables $f(r,\psi)$ defined in polar coordinates inside the region of the unitary circle (in our case shape or electrostatic projection of a surface patch) can be decomposed in the Zernike basis:
\begin{equation}
f(r,\psi)=\sum_{n'=0}^\infty\sum_{m=0}^{n'}c_{n'm}Z_{n'm}(r,\psi),
\end{equation}
where
\begin{equation}
    c_{n'm}=\frac{n'+1}{\pi}\int_0^1dr~r\int_0^{2\pi}d\psi Z_{n'm}^*(r,\psi)f(r,\psi)
\end{equation}
and
\begin{equation}
    Z_{n'm}=R_{n'm}(r)e^{im\psi}.
\end{equation}
$c_{n'm}$ are the expansion coefficients, while the complex functions $Z_{n'm}(r,\psi)$ are the Zernike polynomials. The radial part $R_{n'm}$ is given by
\begin{equation}
    R_{n'm}(r)=\sum_{k=0}^{\frac{n'-m}{2}}\frac{(-1)^k(n'-k)!}{k!\big(\frac{n'+m}{2}-k\big)!\big(\frac{n'-m}{2}-k\big)!}.
\end{equation}

The complete set of Zernike polynomials forms a basis for expanding functions on the unit circle. The set of complex expansion coefficients ${c_{n'm}}$ uniquely reconstructs the original function, with the resolution of this reconstruction dependent on the order of expansion $N=max(n^{'})$. The norms of these expansion coefficients are known as Zernike invariant descriptors, which are invariant to rotations around the origin of the unit circle.\\
To quantify the similarity between two functions in polar coordinates, we compute the Euclidean distance between the corresponding Zernike invariant vectors. For our shape- and electrostatic projections, a smaller distance between the Zernike vectors of two patches indicates higher complementarity (note that one of the patches is inverted in this comparison).\\
In this study, we used$R=9$\AA~and $N=20$, which were identified as the most effective parameters in previous work \cite{Milanetti2021}. Smaller values of $R$ would result in patches that are too small to adequately distinguish interactions between regions. Conversely, a larger radius would encompass non-interacting regions that inherently have low complementarity. The order of expansion $N$ affects the level of detail captured: lower values lead to excessively "smooth" reconstructions, while excessively high $N$ values capture unnecessary details, which can be time-consuming.

\subsection*{Neural network architecture}
The identification of protein-protein interactions relies on learning effective molecular representations. The advent of Deep Learning (DL) methods has significantly enhanced the prediction of molecular properties. These machine learning algorithms use multiple layers to progressively extract higher-level features from raw input data.
In this study, the raw input is derived from the protein structure. The Zernike 2D method provides a compact representation of regions on the protein's molecular and electrostatic potential surfaces, capturing information on shape and electrostatic potential. The scale provided by Di Rienzo \textit{et al.} \cite{DiRienzo2021} allows for a fast evaluation of the hydropathy complementarity.\\
For each residue pair, we generate a $4\times 10$ matrix that represents the normalized complementarity of these three features between the first residue and a 10\AA~radius region centered on the second amino acid.
To build a highly expressive representation from these spatial features, we employ the neural network architecture illustrated in Figure \ref{fig3}b). This network consists of two convolutional layers \cite{krizhevsky2017imagenet} followed by two dense layers. To prevent overfitting, the convolutional layers are followed by a dropout layer for regularization. The learned features are then flattened into a 1D vector and passed through two fully connected layers. The final output layer predicts the probability that the input residue pair is a core interacting pair.

\subsubsection*{Training and hyperparameters}
After selecting 100 complexes for a final blind test, the remaining protein dimers were divided into training (70\%) and testing (30\%) sets. For each protein, a patch was centered on every tenth point on the real interface, and the Zernike formalism was applied to these regions. To ensure a balanced set, an equal number of patches were created from points randomly selected outside the binding regions.\\
For each residue pair, we computed the mean values of the $Z_s$ (shape complementarity) and $Z_{el}$ (electrostatic complementarity) between the patches centered on the respective amino acids. The $H_r$ (hydropathy complementarity) value was also calculated for each pair. Consequently, each residue pair was associated with a matrix and a binary label: 1 if the residues are within 3\AA~from each other and less than 5\AA~from the interface center; otherwise, 0.
The NN was optimized through an Adam version of the stochastic gradient descent \cite{kingma2014adam}.
The training objective was to minimize the binary cross-entropy loss function, comparing the final output prediction to the target label.

\subsubsection*{Blind search for core interacting residues}
To evaluate the accuracy of the NN in a blind search for core interacting residues, we used the 100 complexes that were excluded from the network’s training and testing datasets. For this assessment, we sampled the entire surface of each protein, defining a patch for every set of 10 points. Subsequently, all possible residue pairs were classified by the NN.

\section*{Data Availability}
All relevant data are displayed within the manuscript. Raw data can be requested to the corresponding authors.

\section*{Acknowledgements (not compulsory)}
This research was partially funded by grants from ERC-2019-Synergy Grant (ASTRA, no. 855923), EIC-2022-PathfinderOpen (ivBM-4PAP, no. 101098989), and Project ‘National Center for Gene Therapy and Drugs based on RNA Technology’ (CN00000041) financed by NextGeneration EU PNRR MUR-M4C2-Action 1.4-Call ‘Potenziamento strutture di ricerca e creazione di campioni nazionali di R$\&$S’ (CUP J33C22001130001).

\section*{Author contributions statement}

E.M. and M.M. conceived the work. G.G. designed and trained the neural network, performed analysis and calculations.  L.D.R. curated the dataset. G.R. provided further ideas. All authors interpreted the results, wrote and reviewed the manuscript. 




\bibliographystyle{plain}
\bibliography{main}

\end{document}